\documentclass[manuscript]{aastex}

\slugcomment{\textbf{}}

\shorttitle{2I/Borisov with HST}
\shortauthors{Jewitt et al.}

\begin{document}


\title{Outburst and  Splitting of Interstellar Comet 2I/Borisov}

\author{David Jewitt$^{1,2}$,
Yoonyoung Kim$^3$,
Max Mutchler$^4$,
Harold Weaver$^5$,  
Jessica Agarwal$^{6,3}$
and
Man-To Hui$^{7}$,
}

\affil{$^1$ Department of Earth, Planetary and Space Sciences,
UCLA, Los Angeles, CA 90095-1567\\
$^2$ Department of Physics and Astronomy,
UCLA, Los Angeles, CA 90095-1547\\
$^3$ Max Planck Institute for Solar System Research, 37077 G\"ottingen, Germany\\
$^4$ Space Telescope Science Institute, Baltimore, MD 21218 \\
$^5$ The Johns Hopkins University Applied Physics Laboratory,  Laurel, Maryland 20723 \\
$^6$ Technical University at Braunschweig, Mendelssohnstrasse 3, 38106 Braunschweig, Germany  \\
$^7$Institute for Astronomy, University of Hawaii,  Honolulu, Hawaii 96822\\
}

\email{jewitt@ucla.edu}

\begin{abstract}
We present Hubble Space Telescope observations of a photometric outburst and splitting event in interstellar comet 2I/Borisov.   The outburst, first reported with the comet outbound at $\sim$2.8 AU (Drahus et al.~2020), was caused by the expulsion of solid particles having a combined cross-section $\sim$100 km$^2$ and a mass in 0.1 mm sized particles $\sim 2\times10^7$ kg.  The latter corresponds to $\sim 10^{-4}$ of the mass of the nucleus, taken as a sphere of radius 500 m.  A transient ``double nucleus'' was observed on UT 2020 March 30 (about three weeks after the outburst), having a cross-section $\sim$0.6 km$^2$ and corresponding dust mass $\sim 10^5$ kg.  The secondary was absent in images taken on and before March 28, and in images taken on and after April 03.  The unexpectedly delayed appearance and rapid disappearance of the secondary are consistent with an origin through rotational bursting of one or more large (meter-sized) boulders under the action of outgassing torques, following their ejection from the main nucleus.  Overall, our observations  reveal that the outburst and splitting of the nucleus are minor events involving a negligible fraction of the total mass: 2I/Borisov will survive its passage through the planetary region largely unscathed.

\end{abstract}

\keywords{comets: general --- comets: 2I/2019 Q4 Borisov}

\section{INTRODUCTION}
Comet 2I/Borisov (originally C/2019 Q4 and, hereafter, simply ``2I'') was discovered  
on UT 2019 August 30.  The strongly hyperbolic orbit (eccentricity $e$ = 3.356) convincingly points to  an interstellar origin (Higuchi and Kokubo 2019), presumably beyond the snow-line in the protoplanetary disk of an unknown star.    We do not know for how long 2I has been adrift amongst the stars. Its discovery offers the opportunity to compare and contrast its  properties both with those of solar system comets, and with those of the first-detected interstellar object, 1I/'Oumuamua (``1I'').

Remarkably, 1I and 2I appear quite different. Whereas 1I appeared inert,  2I  closely resembles a typical active solar system comet, with  a prominent and persistent dust coma (Jewitt and Luu 2019, Guzik et al.~2020, Jewitt et al.~2020, Kim et al.~2020, Ye et al.~2020) consisting of non-icy grains (Yang et al.~2020) and showing  spectral lines indicative of on-going sublimation (Fitzsimmons et al.~2019, McKay et al.~2020, Bodewits et al.~2020, Xing et al.~2020).  Both 1I (Micheli et al.~2018) and 2I (Jewitt et al.~2020, Hui et al.~2020) exhibit non-gravitational accelerations, perhaps caused by recoil forces from (curiously, undetected) anisotropic mass loss. Alternative explanations have been advanced for the acceleration of 1I by radiation pressure (Bialy and Loeb 2018, Moro-Martin 2019, Flekk{\o}y et al.~2019).  

Comet 2I passed  perihelion on UT 2019 December 08   at distance $q$ = 2.007 AU.   Hubble Space Telescope (HST) observations taken just before perihelion revealed a small nucleus (effective radius 0.2 $\le r_n \le$ 0.5 km) ejecting predominantly submillimeter sized particles from a large fraction of its surface (Jewitt et al.~2020).  In what follows, we adopt $r_n$ = 0.5 km for our calculations, while acknowledging that this is an upper limit to the nucleus radius.  Additional HST observations bracketing perihelion were used to more completely characterize the comet, finding a mass loss rate in dust $\sim$ 35 kg s$^{-1}$, a nucleus obliquity of $\sim$30\degr, and an isophotal asymmetry interpreted as thermally-lagged emission (Kim et al.~2020).  The small nucleus renders 2I susceptible to torques induced by sublimation mass-loss (Jewitt and Luu 2019, Jewitt et al.~2020).  

A small ($\sim$0.7 magnitudes) optical outburst  was detected in the period UT 2020 March 4 to 9, and tentatively ascribed to a nucleus splitting event (Drahus et al.~2020).  Here we report  short-term changes in the near-nucleus region detected from HST observations obtained before and after this event.

\section{OBSERVATIONS}
Our observations were obtained under the General Observer programs GO 16009, 16041 and 16087. For each program, we used the WFC3 charge-coupled device (CCD) camera, which has pixels  0.04\arcsec~wide, giving  a point-spread function (PSF) with a Nyquist-sampled resolution of about 0.08\arcsec~(corresponding to 145 km at geocentric distance $\Delta$ = 2.5 AU).  We read-out only half of one of the two WFC3 charge-coupled devices (CCDs), giving  an 80\arcsec$\times$80\arcsec~field of view.  The telescope was tracked at the instantaneous non-sidereal rate of the comet (up to about 100\arcsec~hour$^{-1}$) and also dithered to mitigate the effects of bad pixels.  We used the broadband F350LP filter  in order to maximize the throughput and hence the signal-to-noise ratios of the data.  This filter has peak transmission $\sim$ 28\%, an effective wavelength $\lambda_e \sim$ 5846\AA, and a full width at half maximum, FWHM $\sim$ 4758\AA.  In each HST orbit, we obtained six images each of 260 s duration.  The geometric parameters of the observations are summarized in Table (\ref{geometry}).

\section{RESULTS}
\subsection{Morphology} 
\begin{figure}
\plotone{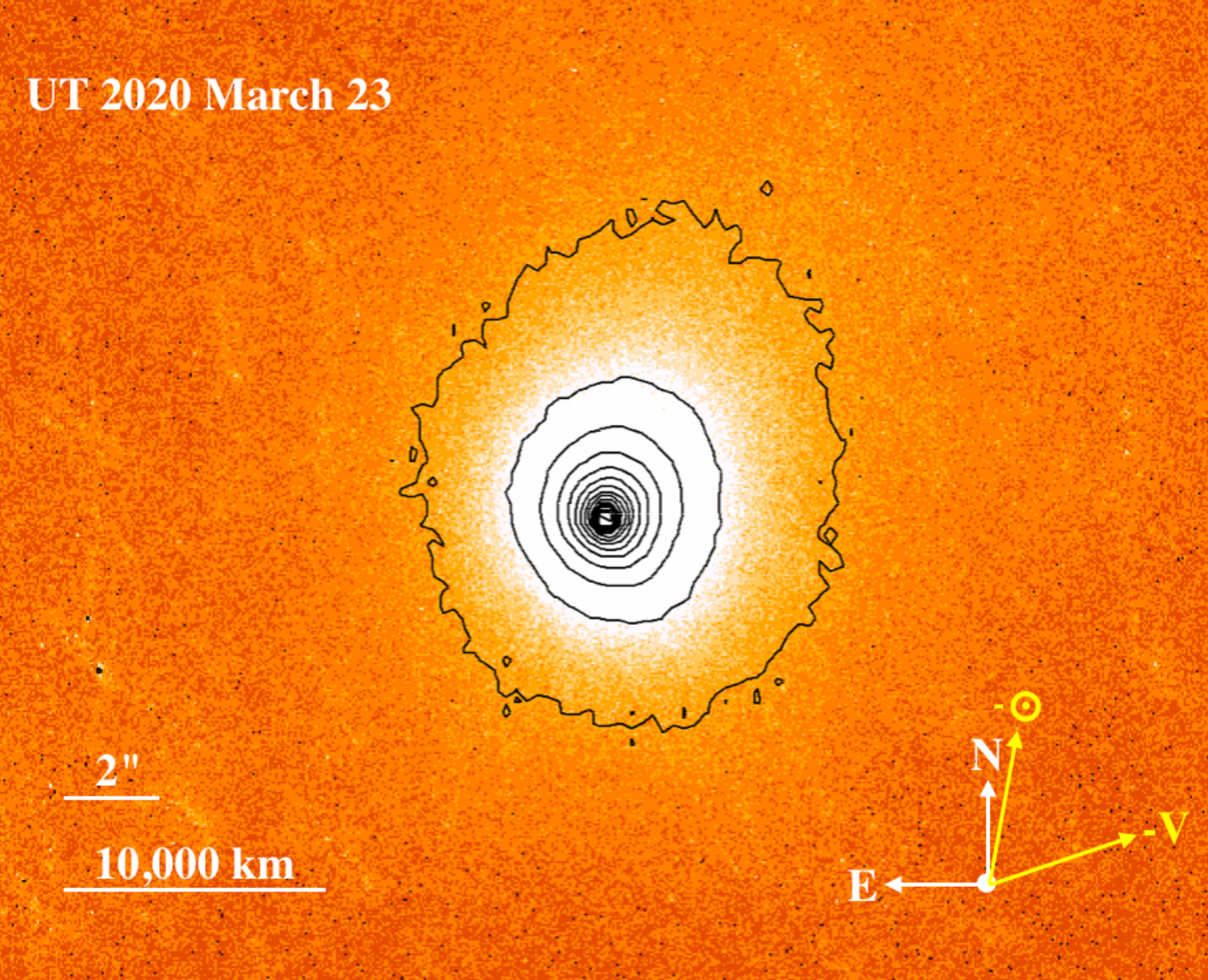}
\caption{Comet 2I/Borisov on UT 2020 March 23, with 2\arcsec~and 10$^4$ km scale bars.     \label{single}}
\end{figure}

A well-established feature of 2I is that the coma dust particles are large, with radii  $a \gtrsim$ 0.1 mm (Jewitt and Luu 2019, Jewitt et al.~2020, Kim et al.~2020), and also slowly-moving, with a characteristic speed for 0.1 mm particles $v \sim$ 9 m s$^{-1}$ (Kim et al.~2020).  As a result, the large-scale morphology of the comet changes relatively little with time and a single-epoch image (Figure \ref{single}) suffices to portray the characteristic rounded coma and stubby  radiation pressure swept tail.  The image was formed by shifting and combining six individual images taken over the $\sim$45 minute HST observing window. We used a clipped median combination (which rejects the highest pixel data number before computing the median of the remainder)  in order to suppress cosmic rays and noise.    Faint structures snaking in the background of the figure are the residuals left by imperfect removal of trailed field stars and galaxies.  While Figure (\ref{single}) closely represents the morphology of 2I in the first six months since its discovery, subtle changes related to the changing observing geometry and, to a lesser extent, to real changes in the comet, do exist (Kim et al.~2020).  

\begin{figure}
\plotone{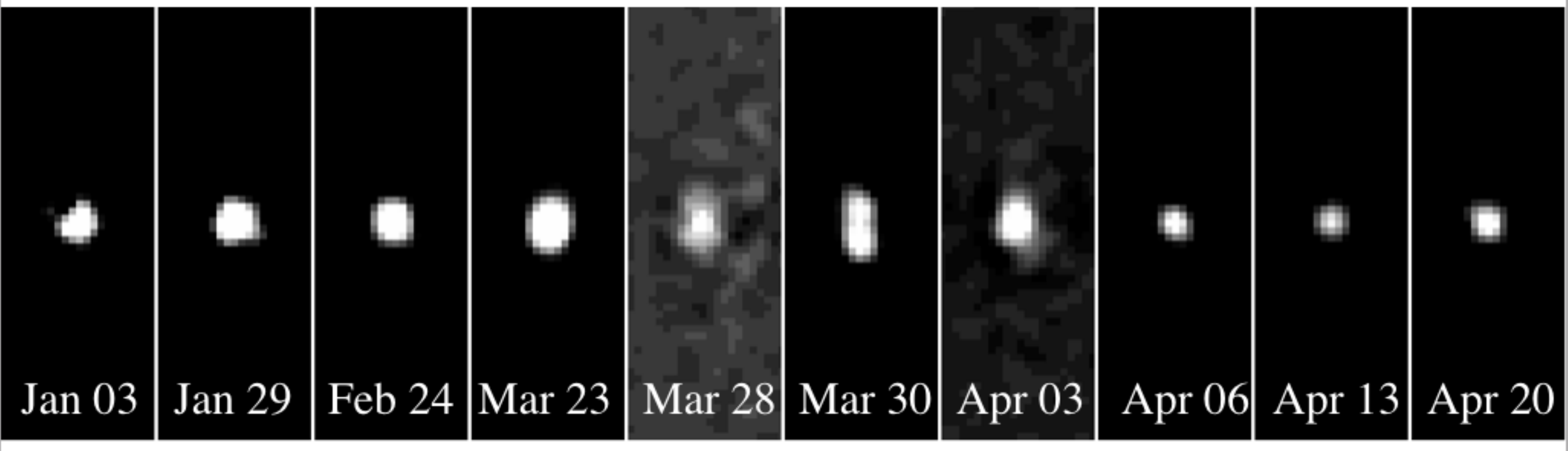}
\caption{Gaussian filtered images from each epoch.  All images are through the F350LP filter except on UT 2020 March 28 and April 03, when narrow filters and/or the use of a polarizer reduce the signal-to-noise ratio relative to the other data.  Each panel has North to the top, East to the left, and is 0.44\arcsec~wide.   \label{gaussed}}
\end{figure}

Here, we focus attention on the near-nucleus morphology of the comet, in particular on changes occurring within a few  $\times10^3$ km of the nucleus.  This central region is shown in Figure (\ref{gaussed}) in the 2020 January ($r_H$ = 2.1 AU) to April ($r_H$ = 3.6 AU) period.  In these images we  suppressed low spatial frequency structures  by subtracting a gaussian-convolved version (standard deviation $\sigma$ = 0.08\arcsec) of the image from  itself. As indicated in Table (\ref{geometry}), most images are directly comparable, because they were taken using a single filter (the broadband F350LP) and followed a consistent observing technique.  We also show an image from March 28 that was obtained under GO 16040 (PI: Bryce Bolin) using several, narrower filters to obtain color information.  We have simply averaged the data taken through different filters in order to maximize the signal-to-noise ratio but, even so, the March 28 image has noticeably lower signal-to-noise compared to the others.   Also in Figure (\ref{gaussed}) the image from April 03 (GO 16044, PI: Qicheng Zhang) employed both different filters and polarizers, while that from UT April 06 (GO 16088, PI: Karen Meech) used F350LP and an exposure sequence more similar to ours.  

The most noteworthy feature of Figure (\ref{gaussed}) is the appearance of a double or ``split'' structure on March 30, with a separation between components $\Delta \theta$ = 0.12\arcsec~(230 km) at a position angle (PA) of $\sim$180\degr.  This structure is evident in the composite and also in the individual (260 s) unprocessed images obtained on this date.  We checked the engineering data to see if a tracking or other error might explain the changed morphology, but found none.  Close inspection of data from the previous visit (March 28), while suffering from a smaller signal-to-noise ratio,  shows marginal evidence for an extension relative to the HST point-spread function, but no evidence for a double.  Another object was reported 0.3\arcsec~north-west from the nucleus in our data from UT March 23 and in the image from UT 2020 March 28 (Bolin et al.~2020) but we are unable to confirm this object on either date.  Data from the subsequent visit (April 03) likewise show a hint of emission to the south west, but this is close to the level of the sky noise and the double appearance is not evident (Zhang et al.~2020).  In short, the core morphology of 2I changed from single to double and reverted back to single within the six day March 28 to April 03 period.  This sudden change stands in stark contrast to the otherwise sedate large-scale morphological development of 2I.

\begin{figure}
\plotone{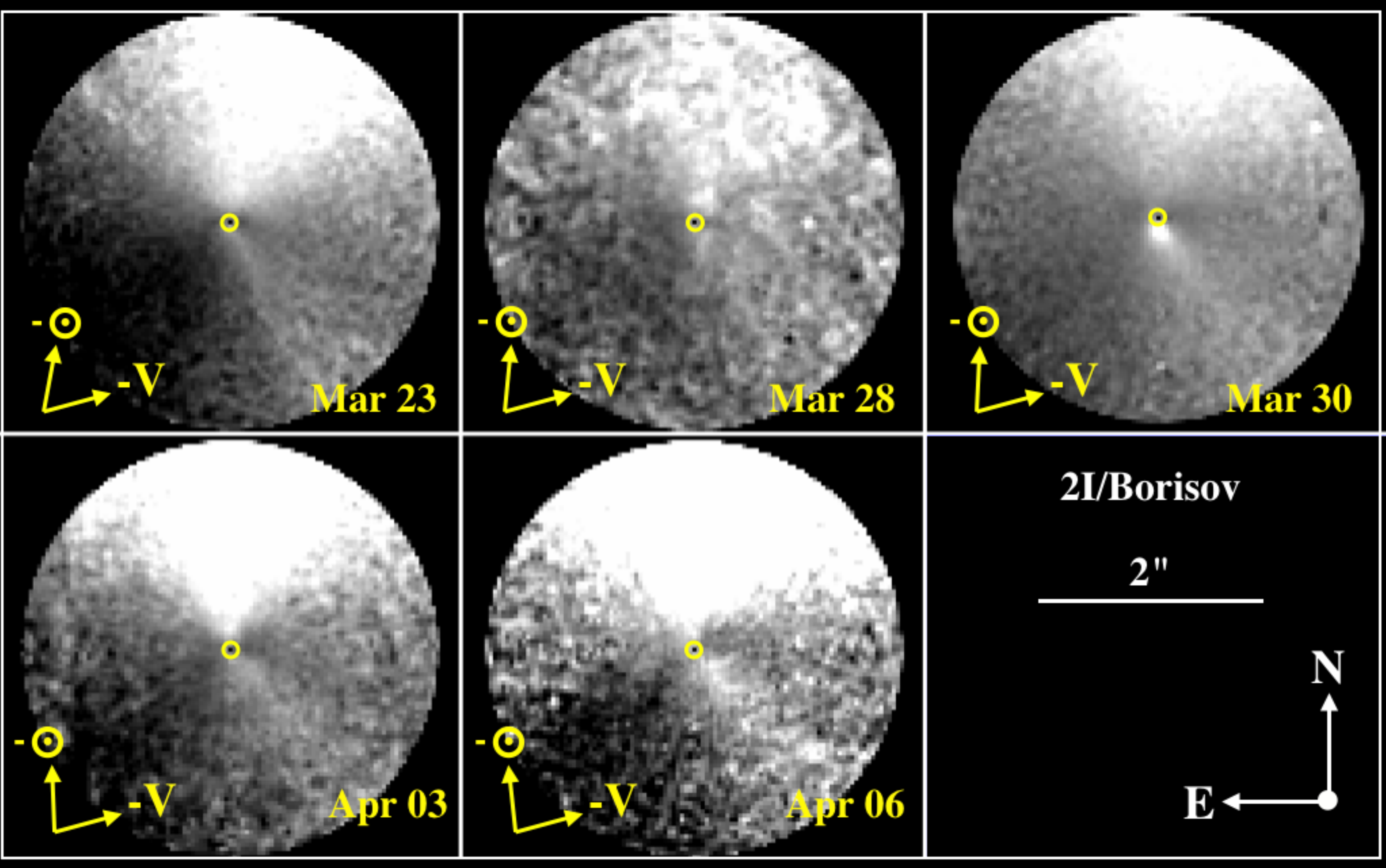}
\caption{Images enhanced by subtracting the median signal computed in a set of concentric annuli, centered on the nucleus, using the algorithm by Samarasinha et al.~(2013).   We focus on the images taken preceding and following the March 30 event.  As in Figure (\ref{gaussed}), all images were taken using the F350LP filter except those from  March 28 and April 03.  Anti-solar and negative velocity vectors are marked by $-\odot$ and $-V$, respectively. \label{enhanced}}
\end{figure}

Figure (\ref{enhanced}) shows the results of a different coma suppression technique, namely the subtraction of the median brightness computed within an annulus centered on the nucleus (Samarasinha et al.~2013).  We used an annulus width of one pixel and used 300 angular sectors.  The annular median subtraction technique enhances azimuthal variations and  suppresses radial variations in the coma surface brightness.  An artifact of the technique is that it systematically suppresses the (circularly symmetric) central object, and hence the nucleus in each panel is gone.  In the case of the March 30 double image, given their closeness, we cannot determine which of the two objects corresponds to the original nucleus and which is the secondary object. We have proceeded on the assumption that the more northerly of the two is the primary body but, since the two components are of similar brightness, our conclusions are not materially affected by this assumption.  The annular median subtraction reveals a persistent fan-shaped coma surface brightness excess to the north (position angles $\sim$350\degr~to 0\degr), coinciding with the direction in which dust particles are blown by solar radiation pressure (see Figure (\ref{single})).  A second, fainter coma excess is evident in position angle $\sim$210\degr. On UT March 30, the secondary body appears as a bright lobe towards the base of this excess.  As is seen more clearly in Figure (\ref{gaussed}), the secondary is separated from the primary by a local brightness minimum and is not simply the root of the  PA 210\degr~jet.

If the boulder was released from the primary at the time of the UT March 4 to 9 photometric outburst (Drahus et al.~2020), the implied sky-plane speed is $V \sim$ 0.13 m s$^{-1}$. This is a lower limit to the true speed  because of the effects of projection.  For comparison, the gravitational escape speed from a non-rotating nucleus is $V_e = (8\pi G \rho/3)^{1/2} r_n$.  With $r_n =$  0.5 km, and density $\rho$ = 500 kg m$^{-3}$ (Groussin et al.~2019) we estimate $V_e =$ 0.26 m s$^{-1}$, broadly consistent with the measured value.  We infer that the boulder ejection was a low energy event, barely capable of launching the body against the gravity of the primary nucleus.  In this regard, 2I resembles solar system comets, in which 0.1 to 10 m s$^{-1}$ ejection velocities are the norm (Sekanina 1997, Boehnhardt 2004).  The Hill radius of 2I, computed for the 2 AU perihelion distance, is $\sim$150 km, precluding the possibility of fallback.

However, given $V \sim$ 0.13 m s$^{-1}$, we should expect to have resolved the double appearance of 2I on March 28 at a separation of 0.11\arcsec~(i.e.~90\% of the separation on UT March 30) but, instead, the second component is absent (Figure \ref{gaussed}).  One possibility is that the boulder was ejected more recently at higher speed.  For example, ejection on March 28 at $V$ = 1.3 m s$^{-1}$ would be needed to reach the 230 km separation by March 30. The boulder would then reach $\sim$680 km (0.34\arcsec)  by April 03, but was not detected.   Therefore, we prefer the interpretation that ejection occurred in early March and that the boulder was present at 0.11\arcsec~on March 28 but intrinsically too faint to detect, then brightened dramatically near March 30, only to fade to invisibility by April 03.

\subsection{Photometry}
We measured photometry within  circular apertures 500, 1,000, 2,000 and 15,000 km in radius when projected to the distance of 2I.  The small apertures sample near-nucleus variations whereas, the 15,000 km aperture is  a better measure of the total light scattered by the comet.
The use of fixed linear (as opposed to angular) apertures provides a measure of the scattering cross-section within a fixed volume surrounding the nucleus, and obviates an otherwise necessary geocentric distance correction that is dependent on the surface brightness distribution.  We obtained background subtraction  from the median signal in a concentric annulus with inner and outer radii 20,000 km and 30,000 km, respectively.  The apparent magnitudes, V$_x$, with $x$ = 500, 1,000, 2,000 and 15,000 are listed in Table (\ref{photometry}), along with absolute magnitudes, H$_x$, computed from 
 
 \begin{equation}
 H_x = V_x - 5\log_{10}(r_H\Delta) - f(\alpha)
 \label{absmag}
 \end{equation}
 
\noindent which corrects for the inverse square law and for the phase function, $f(\alpha)$ at phase angle $\alpha$.  In the absence of a measured value, we assumed $f(\alpha)$ = 0.04 magnitude degree$^{-1}$.  The Table also lists the scattering cross-section, in km$^2$, computed from

\begin{equation}
C_e = \frac{1.5\times10^6}{p_V} 10^{-0.4 H_x}
\label{area}
\end{equation} 

\noindent where $p_V$ is the geometric albedo.  We use $p_V$ = 0.1, as appropriate for  dust in solar system comets (Zubko et al.~2017), while noting that the albedo of this unique interstellar object is likewise unmeasured and could be  significantly different.  The dust albedo is $\sim$2.5 times higher than the average comet nucleus albedo.   
 
\begin{figure}
\epsscale{.80}
\plotone{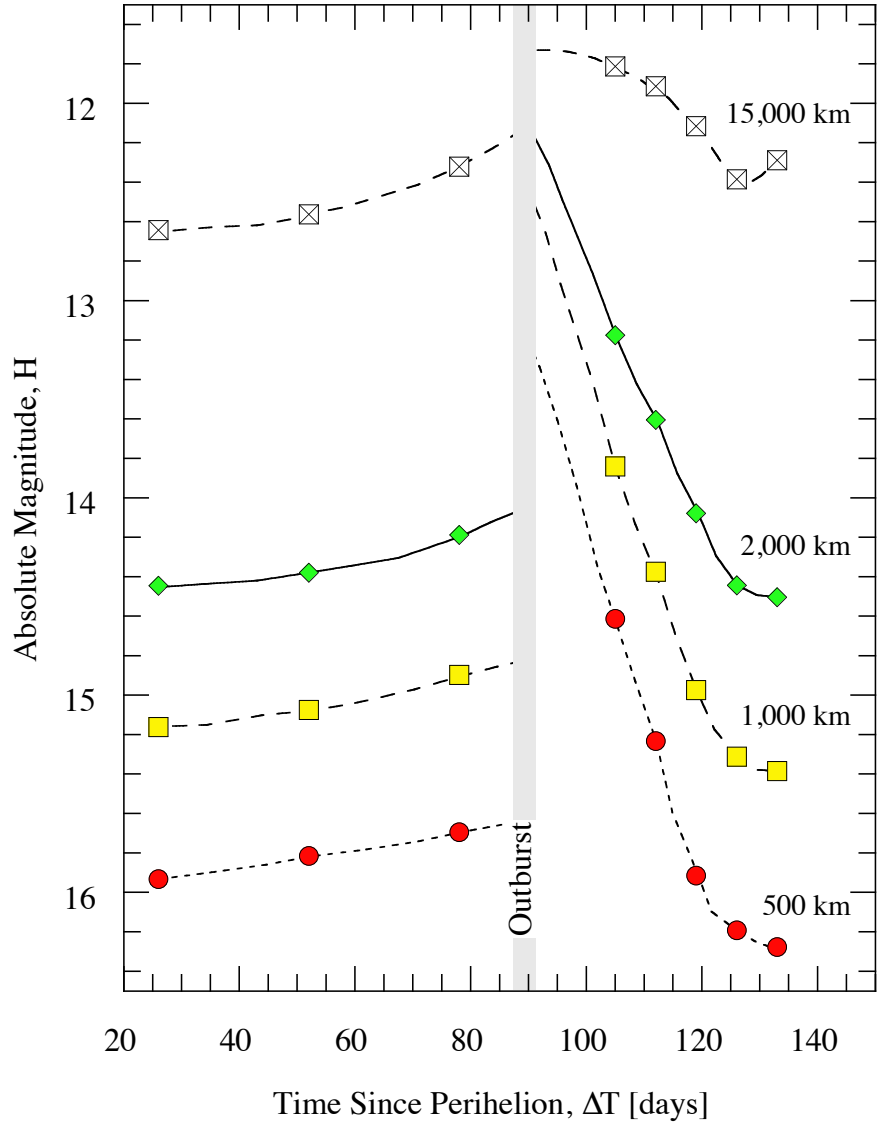}
\caption{Absolute magnitudes vs.~date for the 500 km, 1000 km, 2000 km and 15,000 km radius apertures.   Smoothed dashed-lines are added to guide the eye. The shaded vertical band marks March 4 - 9, the dates of the reported photometric outburst.  \label{absmags_plot}}
\end{figure}

Temporal variations in the absolute magnitudes are plotted in   Figure (\ref{absmags_plot}), in which the UT 2020 March 4.3 to 9.3 dates of photometric outburst (Drahus et al.~2020)  are marked  by a vertical grey band.  Our  data show  that the outburst was preceded by a progressive  brightening of 2I by about 20\%  in all apertures.  This  brightening is not correlated with  the phase angle, which changes modestly and non-monotonically in this period (Table \ref{geometry}), and therefore cannot be ascribed to uncertainties in the phase function.  Neither does it vary with the angle from the orbital plane. Instead, the data show a steady brightening from January 3 to February 24, indicating a period of increasing activity.  We conjecture that this is due to a seasonal effect on the nucleus: according to the pole solution proposed by Kim et al.~(2020), the sub-solar latitude on the nucleus moved from the southern hemisphere to the northern during this period, bringing the Sun's heat to previously unexposed ice.    

Our first post-outburst HST observation occurred about two weeks later (March 23), by which time the near-nucleus brightness had jumped substantially.  The brightening in the 15,000 km ``total light'' aperture increased by $\Delta H \sim$ 0.7 magnitudes (a factor of $\sim$1.9), in agreement with the $\sim$0.7 magnitude brightening in ground-based data using an unstated aperture (Drahus et al.~2020).  The brightening is larger in the smaller apertures, indicative of slowly ejected particles. For example, $\Delta H \sim$ 1.3 magnitudes (a factor of $\sim$3.3) on UT March 23 in the 500 km radius aperture,  falling to 0.8 magnitudes (factor of 2.1) by March 30 and 0.16 magnitudes (factor of 1.2) by April 03.  The sum of the cross-sections of the ejected grains in the 15,000 km aperture measurement increased, relative to the pre-outburst value, by $\Delta C_e \sim$ 100 km$^2$ at the peak on March 23.   This excess disappeared in about 1 month, corresponding to a loss of cross-section at the rate $dC_e/dt =$ -3 km$^2$ day$^{-1}$.
 
 We endeavored to measure the brightness of  the March 30 secondary  in the background-subtracted data from March 30.  In a circle of 0.2\arcsec~radius, with background subtraction from a contiguous annulus extending to 0.8\arcsec, the apparent and absolute magnitudes are V = 23.9 and H = 18.6, respectively, corresponding to a scattering cross-section $C_e$ = 0.6 km$^2$.  The accuracy of this measurement, which refers to the total dust cross-section within the apertue, is both poor and difficult to quantify, being dependent on the size of the residuals left after the removal of the annular median, in an image where the surface brightness gradient is very steep.  We consider $C_e$ = 0.6 km$^2$ as probably no better than a factor-of-two estimate of the cross-section of the encircled dust.
  
 \subsection{Mass}
We estimate the dust mass from the scattered light as follows.  The mass, $M$, of an optically-thin collection of spheres is related to their total cross-section, $C_e$, by $M_e = (4/3) \rho \overline{a} C_e$, where $\overline{a}$ is the cross-section weighted mean radius.  Measurements of 2I show that the particles are large, with estimates from $\overline{a} \sim$ 0.03 to 0.1 mm (Manzini et al.~2020), to $\overline{a} \sim 0.1$ mm (Jewitt and Luu 2019, Jewitt et al.~2020) to $\overline{a} \sim 1$ mm (Kim et al.~2020).  We take $\overline{a}$ = 0.1 mm and $\rho$ = 500 kg m$^{-3}$ to obtain the mass per unit cross-section relation $M_e/C_e = 0.2 $ kg m$^{-2}$.   Therefore, the increase in the cross-section  by $\Delta C_e \sim$ 100 km$^2$, between UT February 24 and March 23 (Table \ref{photometry}), corresponds to a dust mass $\Delta M_e \sim 2\times10^7$ kg.  We take nucleus radius $r_n$ = 0.5 km, the maximum value allowed by high resolution measurements of the surface brightness profile (Jewitt et al.~2020), to estimate nucleus mass $M_n = 3\times10^{11}$ kg.  Then, the fractional mass lost in the outburst is $\Delta M_e/M_n \sim 10^{-4}$.  While photometrically dramatic, the outburst event in 2I is mass-wise inconsequential.  

By similar arguments, the $\sim$0.6 km$^2$ cross-section of the March 30 secondary object corresponds to a dust mass $M \sim 1.2\times10^5$ kg,  equivalent to an equal-density sphere having the modest  radius $\ell \sim$ 3.8 m.  The ratio of the secondary mass to nucleus mass is $M/M_n \sim 4\times10^{-7}$ and the mass contained in the secondary is only 0.6\% of $M_e$.  While a single body of this size fits the data, it is more likely that a considerable number of smaller objects were ejected, forming a  swarm that is unresolved because they are co-located within the projected PSF of the telescope (fragment swarms were also inferred in outbursting comet 17P/Holmes; Stevenson et al.~2010).  The PSF has a width $\sim$ 150 km on UT 2020 March 30.  If so, the total mass would be even smaller than estimated here.  
  
Large fragments are well-known products of the outgassing and decay of solar system comets, but it is not known whether the largest fragments are primordial relics of the accretion process or calved from the nucleus erosively, for example by the collapse of overhangs (Attree et al.~2018).   For example, radar observations of numerous comets reveal abundant particles greater than centimeter size (Harmon et al.~2011), while  observations of bolides associated with cometary meteoroid streams show that larger (meter sized?) objects can be ejected.  Indeed, particles 0.2 m to 2 m in size were directly observed near 103P/Hartley (Kelley et al.~2013, Hermalyn et al.~2013), while ejected particles from 0.1 m to 0.5 m were studied in 67P/Churyumov-Gerasimenko (Davidsson et al.~2015). The record for the latter comet appears to be held by a  $\sim$4 m  boulder detected by  in-situ  imaging\footnote{The data are published, so far, only as a press release;  \url{https://tinyurl.com/yc57ol9z}}.  The nucleus of  67P/Churyumov-Gerasimenko  is strewn with numerous irregularly-shaped, meter-sized boulders (Pajola et al.~2017), deposited on the surface from sub-orbital trajectories.  Fragments released from split comets are routinely $\sim$10s of meters in size (e.g.~Jewitt et al.~2016).  In the case of the small nucleus of 2I, a simple force balance equation (Whipple 1951) indicates that gas drag from the sublimation of water ice can eject 0.2 m size bodies against gravity, while CO gas drag can expel bodies up to 4 m.  These are soft lower limits, however, because gas drag may be aided by centripetal acceleration due to the rotation of the nucleus and, unfortunately, the rotation of 2I is presently not well-established (Bolin 2019).


\section{DISCUSSION}

\subsection{Sublimation}

We use the energy balance equation to calculate  the specific  rate of sublimation of exposed ice, $f_s(T)$ (kg m$^{-2}$ s$^{-1}$),  from

\begin{equation}
\frac{L_{\odot}}{4\pi r_H^2} (1-A) = \chi \left[\epsilon \sigma T(r_H)^4 + H_s f_s(T)\right]
\label{sublimation}
\end{equation}
alp
\noindent Here, the term on the left represents the power absorbed per unit area from the Sun while the terms on the right represent, respectively, the power per unit area radiated from the sublimating surface at temperature, $T$, and the power consumed in sublimating ice at rate $f_s(T)$.  A term to account for thermal conduction has been ignored.  Quantity $L_{\odot} = 4\times10^{26}$ W is the luminosity of the Sun, $A$ is the Bond albedo, $\epsilon$ is the effective emissivity of the surface, $\sigma = 5.67\times10^{-8}$ W m$^{-2}$ K$^{-4}$ is the Stephan Boltzmann constant and $H_s$ is the latent heat of sublimation of the ice.  Few of the parameters in Equation (\ref{sublimation}) are known with confidence, so we make reasonable guesses based on measurements of other comets.  Specifically, we assume  $A$ = 0.04, $\epsilon$ = 0.9 and $\chi$ = 2. Equation (\ref{sublimation}) is then solved iteratively for the temperature and the sublimation rate, $f_s(T)$, using thermodynamic relations for H$_2$O and CO ices from Washburn (1926) and Brown and Ziegler (1980).   

We find from Equation (\ref{sublimation}) that, at $r_H$ = 3 AU (c.f.~Table \ref{geometry}),  $f_s(H_2O) = 2.8\times10^{-5}$ kg m$^{-2}$ s$^{-1}$ and $f_s(CO) = 4.8\times10^{-4}$ kg m$^{-2}$ s$^{-1}$.  If we assume that the nucleus is 0.5 km in radius and in full sublimation from the sunward hemisphere, the equilibrium mass loss rates would be $dM/dt = \pi r_n^2 f_s \sim 22$ kg s$^{-1}$ for H$_2$O, in unreasonably good agreement with water production rates $\sim$20 kg s$^{-1}$  determined from neutral oxygen ([OI]) spectroscopy (McKay et al.~2020).  However, the corresponding equilibrium production of CO is 380 kg s$^{-1}$, about an order of magnitude larger than the measured rate of $\sim$20 kg s$^{-1}$ (Cordiner et al.~2020) to 40 kg s$^{-1}$ (Bodewits et al.~2020). This discrepancy could reflect  sublimation from only a fraction of the surface or reduced sublimation from depths within the nucleus where more nearly interstellar temperatures ($T \lesssim$ 10 K) are preserved.  

Heating by the Sun penetrates a surprisingly small distance into the nucleus, as a result of the small thermal diffusivity, $\kappa \sim$ (1 to 2) 10$^{-9}$ m$^2$ s$^{-1}$, characteristic of porous  material (Sakatani et al.~2018).  While detailed thermophysical calculations are unwarranted, given that the relevant properties of the nucleus are unknown, it is nevertheless informative to make an order of magnitude estimate.  We represent the nucleus of 2I by a  porous dielectric solid with thermal diffusivity $\kappa \sim 10^{-9}$ m$^2$ s$^{-1}$ and  approximate its  inner solar system dynamical (``fly-through'') time as $t_d \sim$ 1 year. Then, from the conduction equation we estimate the e-folding thermal skin depth  $(\kappa t_d)^{1/2} \sim$ 0.2 m. The CO sublimation front would be driven to this or a greater depth from which sublimation would proceed at a temperature and rate lower than calculated from Equation (\ref{sublimation}). 

Recession of the surface due to sublimation occurs at the rate $|d\ell/dt| = f_s/\rho$. With $\rho$ = 500 kg m$^{-3}$, we estimate $d\ell/dt(H_2O) = 6\times10^{-8}$ m s$^{-1}$ and $d\ell/dt(CO) = 1\times10^{-6}$ m s$^{-1}$.  On a one-day timescale, ice thicknesses of $\Delta \ell \sim$ 5 mm (H$_2$O) to 9 cm (CO) could be lost.  In the $\sim$100 days of post-perihelion observation, the nucleus could have lost $\sim$0.5 m by the sublimation of water ice and 9 m by the sublimation of an exposed, pure CO surface (c.f.~Kim et al.~2020).  As noted earlier (Jewitt and Luu 2019, Jewitt et al.~2020), outgassing torques are capable of changing the spin rate of the tiny nucleus of 2I on timescales comparable to the time spent inside the orbit of Jupiter, potentially driving it towards rotational instability. 

\subsection{The Fragment}
The relevant transient characteristics of comet 2I are summarized as:

\begin{enumerate} 
\item The comet underwent an optical outburst between UT 2020 March 4 and 9 (Drahus et al.~2020). 

\item  No secondary was observed until UT 2020 March 30, some 20 days after the  photometric outburst.  The secondary, $\sim$ 230 km sunward from the primary on this date, had disappeared by UT 2020 April 03, only 4 days after it was first observed. 

\item The cross-section, mass and equivalent radius (all dust-dominated) of the secondary were $C_e$ = 0.5 km$^2$,  $M = 3\times10^4$ kg and $\ell \sim$ 3.8 m, respectively.

\end{enumerate}

The outburst likely represents the culmination of a growing period of instability on the nucleus, as indicated by the rising portion of the lightcurve prior to 2020 March in Figure (\ref{absmags_plot}).  While the cause of the outburst cannot be known, several possiblities exist.  Rotational shedding caused by spin-up of the nucleus of 2I is possible (Jewitt and Luu 2019, Jewitt et al.~2020) and future measurements of the rotational period in the absence of coma might test this hypothesis.  Alternatively, the movement of the sub-solar latitude into the northern hemisphere of the nucleus could be implicated, either by causing previously frozen supervolatiles to sublimate, or by inducing thermal stresses capable of triggering landslides or cliff collapse (Steckloff et al.~2016, Pajola et al.~2017) or, perhaps, by triggering the crystallization of amorphous ice (a mechanism suspected in other outbursting comets; Li et al.~2011, Ishiguro et al.~2014, Agarwal et al.~2017).  In the outburst, particulate material occupying a wide range of sizes would be expelled, with most of the $\sim$ 100 km$^2$ increase in scattering cross-section due to smaller particles but with significant mass carried in a few larger bodies.  These larger objects would be independent sources of particles by sublimation but at a low level, commensurate with their individually small cross-sections.  For example, a body with $\ell$ = 1 m would have a cross-section $(r_n/\ell)^{2} \sim 2.5\times10^5$ times smaller than the main nucleus and would sublimate, in equilibrium, at a rate $2.5\times10^5$ times less.  Such a weak secondary source would be optically undetectable, consistent with the non-detections prior to March 30.  However, prolonged outgassing would torque these bodies, driving some to rotational instability.

\subsection{Rotational Bursting}
The sudden appearance of a secondary ``nucleus'' $\sim$20 days after the photometric outburst indicates a delayed instability, rapidly converting the object into finely-divided material having a large total cross-section and, hence, brightness.  Subsequent spreading of the resulting debris cloud  would account for its rapid fading and disappearance only a few days later.  

Sublimation torques  alter the angular momentum and spin rate (angular frequency, $\omega$) of the ejected body.  The natural limit to the spin occurs when a  critical frequency, $\omega_c$, is reached.  At frequencies $\omega \ge$ $\omega_c$  the cohesive strength, $S$, is exceeded by the centripetal force per unit area and the boulder will shed mass.  The process is catastrophic, in the sense that once a boulder fails, the  smaller fragments produced by the breakup have  spin-up timescales even shorter than the original body, and will quickly meet the same fate (Steckloff and Jacobson 2016).  

We describe this instability using a simple model (Jewitt 1997), as follows.
 We approximate the rotationally-induced stress  in a spherical boulder of density $\rho$ and radius $\ell$ by the energy density, $E' = E/V$.  Here, $E$ is the rotational energy and $V_s$ is the volume of the sphere.  We write $E = (1/2) I \omega^2$, where $I = c M \ell^2$ is the moment of inertia and, for a uniform sphere, the constant $c$ = 2/5.  Further substituting $V_s = (4/3)\pi \ell^3$, and setting $E' = S$ as the condition for break-up, we obtain the critical frequency 

\begin{equation}
\omega_c = \left(\frac{5S}{\rho \ell^2}\right)^{1/2}.
\label{omegac}
\end{equation}

A boulder rotating with $\omega \gg \omega_c$ will burst due to its own rotation.  Strength calculations indicate very low cohesive strengths for particulate bodies held together by van der Waals forces, $S \sim$ 25 N m$^{-2}$ (e.g.~Sanchez and Scheeres 2014). Astronomical data from fragmented asteroids provide empirical estimates of $S$ that are of the same order. For example, the fragmented active asteroid P/2013 R3 had $S \sim$ 40 to 210 N m$^{-2}$ (Hirabayashi et al.~2014).  Even smaller tensile strengths, from 1 to 5 N m$^{-2}$, have been deduced from the collapse of overhangs on the nucleus of 67P/Churyumov-Gerasimenko (Attree et al.~2018), although these must refer to the weakest parts of stronger structures.  Substituting representative values $S =$ 10 to 10$^2$ N m$^{-2}$ and $\rho$ = 500 kg m$^{-3}$, we find from Equation (\ref{omegac}) that a boulder of nominal radius $\ell$ = 1 m  would need to spin faster than $\omega_c =$ 0.3 to 1 s$^{-1}$ (i.e.~rotational period $2\pi/\omega_c \sim$ 6 s to 20 s) in order for centripetal forces to exceed cohesion.  

How long would it take for spin-up to $\omega_c$ to occur?  We suppose that boulders ejected from a comet nucleus will initially share the spin of the parent body.  For example, the median rotation period of cometary nuclei is reportedly $\sim$11 hours ($\omega \sim 1.6\times10^{-4}$ s$^{-1}$, Kokotanekova et al.~2017) and we expect that ejected boulders will initially spin at about this rate.  We estimate the timescale for outgassing torques to increase the spin-rate from $\omega$ to $\omega_c$, as follows.    

The reaction force from outgassing is approximated as $F = f_s f_A \pi \ell^2 V_{th}$, where $V_{th}$ is the speed of the sublimated gas and $f_A$ is the fraction of the surface which is outgassing.  We assume $f_A$ = 1 in the following.  The reaction force, in turn, exerts a torque on the nucleus of magnitude $T = k_T F \ell$, where 0 $\le k_T \le 1$ is the dimensionless moment arm ($k_T$ = 0 corresponds to isotropic sublimation from a sphere and $k_T$ = 1 to tangential ejection).  Measurements from 9P/Tempel give 0.005 $\le  k_T \le$ 0.04 (Belton et al.~2011); we take $k_T$ = 0.01 as a middle value. Since torque is just $dL/dt$, where $L$ is the angular momentum, we can estimate the e-folding spin-up timescale, $\tau_s$, from $\tau_s \sim L/T$.  A roughly spherical boulder has $L = I \omega = (4\pi/15)\rho \ell^5 \omega$, from which we estimate

\begin{equation}
\tau_s =  \left(\frac{4\ell}{15 f_s f_A \pi k_T V_{th}}\right) (5 S \rho)^{1/2}
\label{tau_s}
\end{equation}

\begin{figure}
\epsscale{.80}
\plotone{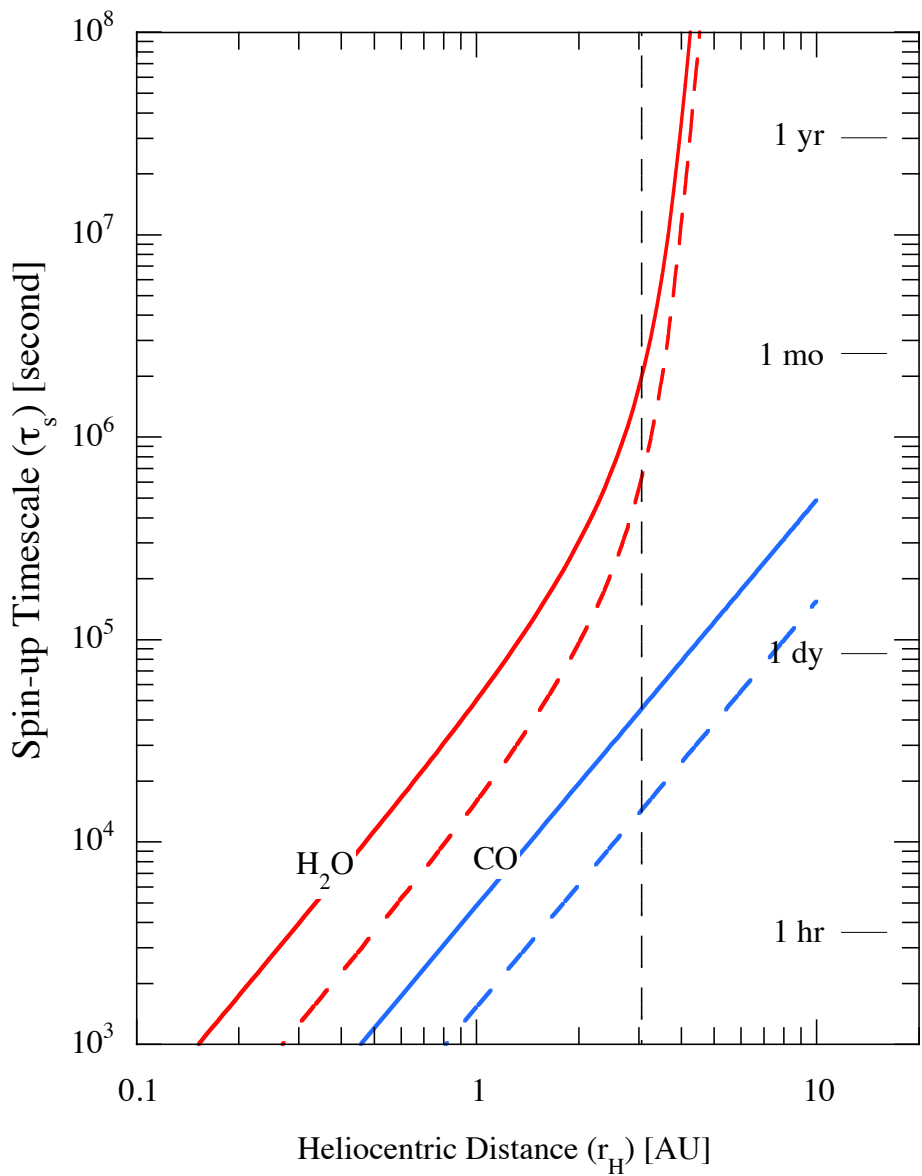}
\caption{Spin-up timescales for the nominal $\ell$ = 1 m radius boulder, computed from Equation (\ref{tau_s}) for sublimating H$_2$O (red lines) and CO (blue lines).  Dashed and solid lines refer to assumed cohesive strengths $S$ = 10 N m$^{-2}$ and $S$ = 100 N m$^{-2}$, respectively.  The  vertical black dashed line marks the heliocentric distance of 2I/Borisov in 2020 March. \label{tau_plot}}
\end{figure}

Solutions to Equation (\ref{tau_s}) are plotted in Figure (\ref{tau_plot}) for  the nominal $\ell$ = 1 m body.  Separate curves are plotted for $f_s$ due to H$_2$O (in red) and CO (in blue) sublimation.  Dashed and solid lines for each volatile refer to  cohesive strengths $S$ = 10 N m$^{-2}$ and $S$ = 100 N m$^{-2}$, respectively.  The blue straight lines (with gradient -2 in log-log space)  for CO reflect the fact that the sublimation term in Equation (\ref{sublimation}) dominates over the radiation term, so that $f_s(CO) \propto r_H^{-2}$.  The figure shows that spin-up times at 3 AU are from hours to $\sim$1 day for CO sublimation and from $\sim$1 week to $\sim$1 month for H$_2$O.  Given the many unknowns and approximations involved, these timescales are clearly no better than order-of-magnitude estimates.  However, they serve to show that the delayed appearance of the secondary on March 30 is consistent with the time needed to spin-up large boulders to rotational bursting (Figure \ref{gaussed}).

Anisotropic reaction forces from outgassing also drive the familiar ``non-gravitational motion'' of comets (e.g.~Marsden et al.~1972).  The magnitude of the non-gravitational acceleration on a boulder, $\zeta$, is obtained from

\begin{equation}
M \zeta = k_R V_{th} \frac{dm}{dt}
\label{accn}
\end{equation}

\noindent where $M$ is the boulder mass, $V_{th}$ is the speed of the escaping material and $dm/dt$ is the mass loss rate.  The dimensionless quantity $k_R$ is the fraction of the escaping momentum delivered to linear motion nucleus, with $k_R$ = 0 corresponding to isotropic mass loss and $k_R$ = 1 corresponding to perfectly collimated emission.  In time, $t$, a constant acceleration $\zeta$ should propel the fragment over a distance, $x = \zeta t^2/2$.  Assuming a spherical boulder of radius $\ell$ and substituting into Equation (\ref{accn}), we find 

\begin{equation}
k_R = \frac{8\rho x}{3 f_s V_{th} t^2} \ell
\end{equation}

\noindent We set $t$ = 20 days ($1.7\times10^6$ s), equal to the lag between the photometric outburst and the appearance of the fragment.  The observed separation, $x$ = 230 km, is a lower limit to the true separation because of the effects of projection. Substituting, we obtain $k_R/\ell \gtrsim$ 0.01.  For a 1 m fragment, the measured separation implies $k_R \sim$  0.01, for a 4 m fragment, $k_R \sim$ 0.04.   This is about an order of magnitude smaller than $k_R$ estimated for the well-characterized nucleus of 67P/Churyumov-Gerasimenko (Appendix).  

Small values of $k_R$ are a natural consequence of rapid rotation, particularly when the rotation period is shorter than the cooling time for the body.  In this case, heat deposited on the day side of the boulder is  transported to the night side before it can be lost by radiation or latent heat effects, leading to a latitude-isothermal temperature distribution and the suppression of the day-night temperature asymmetry.  In turn, this will decrease $k_R$ relative to its value in a slowly-rotating nucleus like that of 67P.  On the other hand, while suppressing $k_R$, rapid rotation should have no effect on the torque, since the dimensionless moment arm, $k_T$,  is  mainly a function of the shape of the body, not the day-night temperature asymmetry.  The critical period for the onset of rotational averaging of the surface temperature, $\tau_c \sim$ 1 hour, is derived in the Appendix.

The same process should operate in other comets, and may be responsible for the delayed appearance and rapid disappearance of fragments observed in split comets.  For example, about a third of the fragments in comet 332P/Ikeya-Murakami appeared between the first and the third of three consecutive days of observation (UT 2016 January 26, 27 and 28; Jewitt et al.~2016) and delayed from their ejection by about two months.  Spacecraft observations of large secondaries (e.g.~Agarwal et al~2016, Fulle et al.~2016) typically sample the near-nucleus space, giving insufficient time for spin-up disruption to occur.  Furthermore, we are aware of no reports of independent sublimation of these fragments suggesting that ejected bodies may  be less icy than in the case of 2I. 

Once disrupted, the particles released by rotational instability of a boulder would leave its surface at the equatorial velocity, $v = \ell \omega_c$ which, by substitution into Equation (\ref{omegac}), is $V \sim$ 0.5 m s$^{-1}$ (independent of $\ell$).  The particle cloud created by the fragmented boulder will appear optically thick for a time $t_1 \sim (C_e/\pi)^{1/2}/V$, where $C_e$ is the sum of the cross-sections of the fragments.   For example, with $C_e$ = 0.6 km$^2$, $t_1 \sim 10^3$  s.  For times $t \le t_1$, the debris cloud will appear as an expanding sphere, with cross-section and scattered light increasing in proportion to $t^2$.  Timescale $t_1$ is so short that it is unlikely to be observed.  Once optically thin, the debris cloud from a fragmented boulder will remain unresolved until cloud radius exceeds the size of the projected PSF.  In the case of the present data, the  0.04\arcsec~radius of the PSF corresponds to $\sim$ 73 km at 
$\Delta$ = 2.5 AU, giving a crossing time $t_f \sim1.5\times10^5$ s (about 1.7 day).  At  times, $t \ge t_f$, the cloud will become resolved and the core will fade as the debris moves out of the PSF.  This crudely-estimated fading time ($\sim$days) is consistent with the sudden disappearance of the fragment between March 30 and April 03 (Figure \ref{gaussed}).

Lastly, Figure (\ref{tau_s}) shows that transient secondaries caused by rotational bursting can be expected only within a narrow range of heliocentric distances.  Close to the Sun ($r_H \lesssim$ 1 AU), the spin-up timescales are so short that  fragments ejected with characteristic meter per second speeds cannot be resolved from the parent nucleus, even in HST data.  At best, the rotational fragmentation of secondaries in near-Sun comets might be inferred from  impulsive brightening events in the lightcurve.  Far from the Sun ($r_h \gtrsim$ 4 AU), the water ice sublimation spin-up timescale  exceeds the timescale for the change of heliocentric distance ($r_H/(dr_H/dt) \sim 1$ yr, for a comet in free-fall at 4 AU).  In this case, the outgassing torque  falls towards zero before the critical breakup frequency  is reached.    Our observations of 2I fall neatly within this range of distances.

\clearpage 

\section{SUMMARY}
We present high resolution observations of the interstellar comet 2I/(2019 Q4) Borisov, both before and after the UT 2020 March 4 to 9 photometric outburst (Drahus et al.~2020). The data reveal the delayed appearance of a short-lived, low-mass fragment. Collectively, the data show that the outburst and fragmentation of the nucleus were minor events.

\begin{enumerate}

\item The outburst released material with a combined cross-section $\sim$100 km$^2$ (geometric albedo 0.1, assumed), and an estimated mass $\sim 2\times10^7$ kg, equal to about 10$^{-4}$ that of the nucleus.

\item  A transient double nucleus was observed on  UT 2020 March 30, about three weeks after the photometric outburst, with a  separation  between components  $\sim$0.12\arcsec~(230 km at the comet) in position angle $\sim$180\degr. The second component was not observed on March 28 or April 03.

\item The cross-section of the secondary object, $\sim$0.6 km$^2$ (geometric albedo 0.1, assumed), corresponds to a mass of 0.1 mm particles $M \sim 1.2\times10^5$ kg.  The radius of a single,  equal mass sphere of density $\rho$ = 500 kg m$^{-3}$ is $\ell$ = 3.8 m.  However, the secondary likely consists of an unresolved collection of a larger number of smaller boulders.

\item The delayed appearance and rapid demise of the secondary together suggest an origin by spin-up and rotational bursting of one or more large (meter-scale) boulders under the action of outgassing torques.

\end{enumerate}

\acknowledgments

We thank the anonymous referee for comments.  This work was supported under Space Telescope Science Institute programs GO 16009, 16041 and 16087.  Y.K. and J.A. were supported by the European Research Council (ERC) Starting Grant No. 757390 (CAstRA).

{\it Facilities:}  \facility{HST}.

%
%
%
%

\clearpage

\begin{deluxetable}{llccrcccrr}
\tabletypesize{\scriptsize}
\tablecaption{Observing Geometry 
\label{geometry}}
\tablewidth{0pt}
\tablehead{ \colhead{UT Date and Time}   & GO\tablenotemark{a} & FLTR\tablenotemark{b} & $\Delta T$\tablenotemark{c} & \colhead{$r_H$\tablenotemark{d}}  & \colhead{$\Delta$\tablenotemark{e}} & \colhead{$\alpha$\tablenotemark{f}}   & \colhead{$\theta_{\odot}$\tablenotemark{g}} &   \colhead{$\theta_{-V}$\tablenotemark{h}}  & \colhead{$\delta_{\oplus}$\tablenotemark{i}}   }
\startdata


2020 Jan 03 03:22 - 03:57 		& 16009 & F350LP & 26 & 2.086 & 1.942 & 23.0 & 294.5 & 319,4 & -9.0 \\
2020 Jan 29  11:35 - 12:08 	& 16041 & F350LP &52 & 2.313 & 2.059 & 25.2 & 305.0 & 304.3 & 0.2 \\
2020 Feb 24 02:42 - 03:19 	& 16041 & F350LP &78 & 2.638 & 2.269 & 21.6 & 321.9 & 2902 & 7.8 \\
2020 Mar 23 10:17 - 10:54 		& 16041 & F350LP &105 & 3.072 & 2.565 & 17.6 & 349.9 & 285.6 & 12.8 \\
2020 Mar 28 01:19 - 03:31 		& 16040 & Various\tablenotemark{j}  & 110 & 3.148 & 2.620 & 17.0 & 355.2 & 286.0 & 13.2 \\
2020 Mar 30 06:00 - 06:37 		& 16087 & F350LP &112 & 3.184 & 2.646 & 16.7 & 357.8 & 286.3 & 13.3 \\
2020 Apr 03 03:14 - 04:32 		& 16044 & F606W & 116 & 3.250 & 2.695 & 16.2 & 2.4 & 286.8 & 13.6 \\
2020 Apr 06 00:34 - 01:10			& 16088 & F350LP & 119 & 3.298 & 2.731 & 15.8 & 5.9 & 287.3 & 13.7 \\
2020 Apr 13 16:53 - 17:34 		& 16087 & F350LP &126 & 3.310 & 2.740 & 15.7 & 6.7 & 287.4 & 13.7 \\ 
2020 Apr 20 15:43 - 16:13 		& 16087 & F350LP & 133 & 3.552 & 2.932 & 14.1 & 23.8 & 289.4 & 13.7 \\

\enddata


\tablenotetext{a}{HST General Observer Program Number}
\tablenotetext{b}{Filter employed}

\tablenotetext{c}{Number of days from perihelion (UT 2019-Dec-08). }
\tablenotetext{d}{Heliocentric distance, in AU}
\tablenotetext{e}{Geocentric distance, in AU}
\tablenotetext{f}{Phase angle, in degrees}
\tablenotetext{g}{Position angle of the projected anti-Solar direction, in degrees}
\tablenotetext{h}{Position angle of the projected negative heliocentric velocity vector, in degrees}
\tablenotetext{i}{Angle of Earth above the orbital plane, in degrees}
\tablenotetext{j}{F438W, F689M, F845M}

\end{deluxetable}

\clearpage

\begin{deluxetable}{lccccc}
\tabletypesize{\scriptsize}

\tablecaption{HST Fixed-Aperture F350LP Photometry\tablenotemark{a} 
\label{photometry}}
\tablewidth{0pt}

\tablehead{ \colhead{UT Date} & $\Delta T$ & $\ell$ = 500 km & 1,000 km  & 2,000 km & 15,000 km}
\startdata
2020 Jan 03 & 26 	& 19.89/15.93/6.36 	& 19.12/15.16/12.9 	& 18.40/14.45/25.0 & 16.60/12.64/132\\
2020 Jan 29 & 52 	& 20.21/15.82/7.08 	& 19.47/15.08/14.0 	& 18.78/14.38/26.6 & 16.96/12.56/141\\
2020 Feb 24 & 78 	& 20.44/15.69/7.91 	& 19.65/14.90/16.5 	& 18.94/14.19/31.7 & 17.07/12.32/177\\
2020 Mar 23 & 105 	& 19.80/14.61/21.4 	& 19.03/13.84/43.6 	& 18.36/13.18/80.5 & 17.00/11.81/282\\
2020 Mar 30 & 112 	& 20.53/15.23/12.1 	& 19.67/14.389/26.7 	& 18.90/13.61/54.2 & 17.21/11.91/257\\
2020 Apr 06 & 119 	& 21.32/15.91/6.46 	& 20.38/14.97/15.4 	& 19.48/14.08/35.0 & 17.52/12.12/214\\
2020 Apr 13 & 126 	& 21.61/16.19/5.00 	& 20.73/15.31/11.3 	& 19.86/14.44/25.1 & 17.80/12.38/167\\
2020 Apr 20 & 133 	& 21.93/16.28/4.63 	& 21.04/15.39/10.5 	& 20.16/14.50/23.7 &17.94/12.29/182 \\

\enddata

\tablenotetext{a}{For each date and aperture radius, $\ell$, the Table lists the apparent magnitude, V, the absolute magnitude, H, and the scattering crossection, $C_e$ (in units of km$^2$), in the order V/H/$C_e$. H and $C_e$ are computed  using Equations (\ref{absmag}) and (\ref{area}).}

\end{deluxetable}

\clearpage

\clearpage

\clearpage

\clearpage

%
%

\clearpage

\clearpage

\clearpage

\appendix
\section{Momentum transfer coefficient, $k_R$}
First, we estimate the momentum transfer coefficient, $k_R$, using data for the best-characterized cometary nucleus, 67P/Churyumov-Gerasimenko.  The 67P nucleus mass is $M_n = 1.0\times10^{13}$ kg (Patzold et al.~2016).  The  perihelion production rate is dominated by water at $Q_{H_2O} = 1\times 10^{28}$ s$^{-1}$, corresponding to $dm/dt$ = 300 kg s$^{-1}$, while the outflow speed is measured at $V_{th}$ = 0.9 km s$^{-1}$ (Biver et al.~2019). The non-gravitational acceleration parameters of 67P are listed on the JPL Horizons site\footnote{\url{https://ssd.jpl.nasa.gov/horizons.cgi}} as A$_1$, A$_2$, A$_3$ = 1.1$\times10^{-9}$, -3.7$\times10^{-11}$, 2.5$\times10^{-10}$ AU day$^{-2}$.  We compute the total acceleration of the nucleus from 

\begin{equation}
\zeta = g(r_H) (A_1^2 + A_2^2 + A_3^2)^{1/2}
\label{A1}
\end{equation}

\noindent where $g(r_H)$ is the dimensionless function defined by Marsden et al.~(1972).  Evaluated at perihelion we find $g(1.24)$ = 0.61, giving  $\zeta = 1.3\times10^{-8}$ m s$^{-2}$ by Equation (\ref{A1}).  Equation (\ref{accn}) then gives

\begin{equation}
k_R = \frac{M_n\zeta}{V_{th} (dm/dt)}
\end{equation}

\noindent from which we evaluate $k_R$ = 0.5.

Second, we estimate the critical rotation period of a body below which rotational averaging of the temperature becomes important. We compare the heat content per unit area of the day side, $H = \rho \delta c_p T$, with the rate of loss of heat per unit area, $dH/dt$.    Here, $c_p$ is the specific heat capacity and $\delta$ is the thickness of the diurnally heated layer of the body, given rotation period, $\tau_c$.  Neglecting numerical factors of order unity, we write $\delta \sim (\kappa \tau_c)^{1/2}$, with $\kappa$ being the thermal diffusivity.  For simplicity, we assume that the heat is lost entirely by sublimation (a good approximation at small $r_H$), so $dH/dt = f_s H_s$, and use Equation (\ref{sublimation}) to calculate $f_s$, as above.   Then the  e-folding timescale for losing daytime heat is $\tau_c = H/(dH/dt)$, or

\begin{equation}
\tau_c = \frac{ \rho \delta c_p T}{f_s H_s}.
\end{equation}

\noindent We substitute for $\delta$ and solve to find


\begin{equation}
\tau_c = \kappa \left(\frac{\rho c_p T}{f_s H_s}\right)^2.
\label{spinner}
\end{equation}

\noindent For H$_2$O ($H_s = 2\times10^6$ J kg$^{-1}$), solution of Equation (\ref{sublimation}) gives $f_s = 2.8\times10^{-5}$ kg m$^{-2}$ s$^{-1}$ while for CO ($H_s = 2\times10^5$ J kg$^{-1}$), $f_s = 4.8\times10^{-4}$ kg m$^{-2}$ s$^{-1}$, so that the product in the denominator of Equation (\ref{spinner}), $f_s H_s$, does not differ much between the two volatiles.  For water sublimation at $r_H$ = 3 AU, solution of Equation (\ref{sublimation}) gives $T$ = 150 K.  With $\kappa = 10^{-3}$ W m$^{-1}$ K$^{-1}$, $\rho = 500$ kg m$^{-3}$ and $c_p = 10^3$ J kg$^{-1}$ K$^{-1}$, Equation (\ref{spinner}) gives $\tau_c \sim$ 2000 s, or about 1 hour. Bodies with rotation periods $<\tau_c$ will suffer rotational averaging of their  day-night temperature contrast, reducing both $k_R$ and the magnitude of the non-gravitational acceleration (Equation \ref{accn}) and shrinking the distance traveled relative to more slowly rotating bodies. A boulder spun-up by outgassing torques will quickly find itself in this latitude-isothermal, small acceleration regime.  


%

\end{document}